\documentclass[12pt]{iopart}

\usepackage{amsfonts}
\newcommand{\ket}[1]{\mbox{$| #1 \rangle$}}
\newcommand{\bra}[1]{\mbox{$\langle #1 |$}}

\newcommand{\braket}[2]{\mbox{$\langle #1 | #2 \rangle$}}

\begin{document}

\title{The Gross-Pitaevskii Equation and Bose-Einstein condensates}

\author{J. Rogel-Salazar}

\address{Applied Mathematics and Quantitative Analysis Group,\\
  Science and Technology Research
  Institute,\\ School of Physics Astronomy and Mathematics,\\
  University of Hertfordshire, Hatfield, AL10 9AB, U.K.}
\ead{j.rogel@physics.org}
\begin{abstract}
  The Gross-Pitaevskii equation is discussed at the level of an
  advanced course on statistical physics. In the standard literature the
  Gross-Pitaevskii equation is usually obtained in the framework of
  the second quantisation formalism, which in many cases goes beyond
  the material covered in many advanced undergraduate courses. In this
  paper, we motivate the derivation of the Gross-Pitaevskii equation
  (GPE) in relationship to concepts from statistical physics,
  highlighting a number of applications from dynamics of a
  Bose-Einstein condensate to the excitations of the gas cloud.  This
  paper may be helpful not only in encouraging the 
  discussion of modern developments in a statistical mechanics course,
  but also can be of use in other contexts such as mathematical
  physics and modelling. The paper is suitable for undergraduate and
  graduate students, as well as general physicists.

\end{abstract}

% Uncomment for PACS numbers title message
\pacs{01.40.Fk, 02.30.Xx, 05.30.Jp, 03.75.Hh, 03.75.Mn, 67.10.-j}
\vspace{2pc}
\noindent{\it Subjects}: Quantum Gases, Statistical Physics,
Education, Bose-Einstein Condensation \\
\submitto{\EJP}
\maketitle

\section{Introduction}
\label{ch:intro}
The realisation of Bose-condensed gases with alkali elements
\cite{Anderson95, Davis95} provided physicists with a great
opportunity to test a new regime of matter that until then was
considered a purely theoretical concept. The theoretical basis for the
description of these systems has seen the development of excellent
reviews \cite{Dalfovo99, Leggett01} and textbooks \cite{Pethick_Book02,
  Griffin_Book09}. The dynamics of the condensate at zero temperature
is generally described with the Gross-Pitaevskii equation (GPE) which
is effectively a mean-field approximation for the interparticle
interactions. It is usually the case that the GPE is derived from the
framework of second quantisation, which may pose some barriers to
exposing the description of trapped Bose-condensed gases to newcomers
to the field. Since Bose-Einstein condensation (BEC) can be understood
from statistical correlations, without the need of boson-boson
interaction, there is a feasible possibility of discussing the
phenomenon beyond the usual Bose-Einstein statistics
\cite{Ferrari10}. With that in mind, the aim of this paper is to
enable students and lecturers to approach the description of a BEC in
an accessible manner which allows them to tackle more advanced
treatments using concepts readily available to them.

\section{Bose-Einstein condensation}
\label{subsec:BEC}
From the description provided by thermodynamics, within a gas, all the
particles behave in the same manner and in principle they can occupy
certain quantum states, that is certain energy states
\cite{Gasser_Book1995}. If these particles are {\it fermions}, two or
more of them cannot occupy the same quantum state (i.e. the Pauli
exclusion principle).  However, if they are {\it bosons}, any number
of them can occupy the same quantum state. When we put these particles
in a given configuration, they will get distributed in the energy
levels of such a configuration, with an increasing occupation of the
states of minimum energy as we lower the temperature.

For a collection of bosons, and in the limit where the temperature
goes to zero, all the particles are going to occupy the ground state
of the system.  Actually, before achieving this limit, an spectacular
accumulation of particles in the ground state can be observed.
Therefore, for a sufficiently low temperature, the majority of the
particles are in the same quantum state, and have the same velocity.
In this way, the collection of bosons behaves like a macroscopic fluid
with new properties, such as superfluidity \cite{Landau_Book1980}.

In order to study these properties, it is only necessary to
concentrate on
the ground state. The state of the particles in a certain level is
described by a normalised function, which is an eigenvector of the
Hamiltonian associated to the physical situation under consideration,
and whose eigenvalue corresponds to the energy of that level. Thus, we
have to find the eigenvectors with minimum eigenvalue for the
Hamiltonian that describes our system. This Hamiltonian includes all
the interactions between every pair of particles. In other words, if
there are $N$ particles in the system, the Hamiltonian will have $N^2$
terms. In order to simplify the problem, we make use of the
``mean-field'' approximation, which means that the action felt by a
given particle due to the rest is substituted by the mean action of
the fluid over the particle.  This approximation is good if one can
neglect the correlations in the gas, that is, if the gas is diluted,
which is the case for the condensates obtained experimentally. In this
way, we have changed a complex model for the interaction among bosons
by a very simplified one that is valid for diluted gases. We now have to
find now an approximate value for the first eigenvalue of this new
Hamiltonian by minimising a functional within a restricted class of
Ansatz - the functions for which all the bosons are in the same state.

As we mentioned above, the interactions between bosons are not
necessary for condensation to take place \cite{Ferrari10}; however,
they play a very important role in the properties of the condensate as
we will see later on. In that sense, using the usual Schr\"odinger
equation \cite{Gasser_Book1995} is not enough; instead we require the
use of the Gross-Pitaevskii equation (GPE).  Other applications of the
GPE include the modelling of superconductors \cite{DePalo1999}
(superconduction is a quantum mechanical effect manifested at low
temperatures and it could be interpreted as some sort of superfluidity
of electron pairs), also to describe optical vortices
\cite{Rozas1997}, that resemble small twisters in a superfluid.

\subsection{A model for superfluids and BECs}
\label{subsec:model}

When helium gas (isotope $^4$He) is cooled down, it is possible to
observe a transition into a liquid phase at a temperature of 4.2 K and
at a pressure of 1 atmosphere. If we continue cooling down the system
well under 2.17 K, this liquid phase acquires highly unusual
properties: it becomes a superfluid \cite{Nozieres_Book1990}.

Superfluidity is manifested, among other effects, by the lack of
viscosity; in other words, the liquid flows without friction. Thus, if
we launch some of this superfluid helium into a ring-shaped channel
for example,
it will not stop. If we try to
move an object up to a certain velocity across the surface of the
liquid, it will not experience any resistance. Another interesting
feature of the system is the 
creation of vortices, which can be seen as small twisters inside the
liquid, that behave quite differently from the ones observed in water
for example. These vortices have certain quantisation properties (the
velocity of the liquid cannot take any arbitrary value). 

These properties, observed in liquid helium $^4$He in 1937, were also
seen in helium $^3$He at a lower temperature and more recently in
Bose-Einstein condensates \cite{Balibar07}. As discussed above, a BEC
is a particular state of matter achieved at ultra cold
temperatures. These phenomena are the macroscopic manifestation of
quantum effects, and to study them it is therefore required to use
quantum mechanics.

In the 1950s, Landau and Ginzburg proposed to model the electrons that
give rise to superconductivity as a superfluid with the aid of an 
equation of the simplified form \cite{Ginzburg50}
\begin{equation}
  \label{eq:LGE}
  i\hbar\frac{\partial u}{\partial t}=-\frac{\hbar^2}{2}\nabla^2 u+g u(|u|^2-1),
\end{equation}
in the frame of a two-fluid model, where $|u|^2$ represents the
superfluid density, which flows without friction, whereas the rest of
the matter is supposed to be in a normal fluid state, and $g$ is a
proportionality constant with dimensions of energy. This equation
renders satisfactory predictions, but its usage is limited due to the
fact that the interactions in a liquid such as helium are fairly
strong.

However, the Landau-Ginzburg model is a particular case of an equation
that emerges quite naturally when we study the behaviour of
Bose-Einstein condensates up to a first order approximation. This
equation is known as the Gross-Pitaevskii equation (GPE)
\begin{equation}
  i\hbar\frac{\partial \psi}{\partial
    t}=\left(-\frac{\hbar^2}{2m}\nabla^2+V_{ext}+g|\psi|^2\right)\psi, 
  \label{eq:GPE}
\end{equation}
where $m$ is the mass of the atoms of the condensate, $|\psi|^2$ is
the atomic density, $V_{ext}$ represents an external potential and $g$
is a parameter that measures the atomic interactions. The GPE has the
same mathematical form as the nonlinear Schr\"odinger equation (NLSE),
which is basically the Schr\"odinger equation
\begin{equation}
  i\hbar\frac{\partial \psi}{\partial
    t}=\left(-\frac{\hbar^2}{2m}\nabla^2+V_{ext}\right)\psi, 
  \label{eq:Schroed}
\end{equation}
plus a nonlinear term that in this case takes into account the interaction between
the particles. In the case of Bose-Eintein condensates, the interactions are so
weak that the predictions made with this equation are very good.

In the following section we will derive the GPE and will use it to
present some of the problems that have been studied within the
mean-field approximation and which can be used as examples to follow up
discussions in undergraduate and graduate lectures related to this
subject such as statistical physics, quantum mechanics or mathematical
physics. We will study the simple case of a fluid in one
dimension, and we will show that if the bosons have an attractive
scattering length, the fluid is no longer stable for certain sizes of
the system. We will also address the excitations in the Bose-condensed
gas. 

\section{The Gross-Pitaevskii equation}
\label{sec:GPE}
As we have discussed above, a Bose-Einstein condensate is obtained
from a collection of bosons in the ground state at very low
temperatures. We can therefore ask 
about the energy of the ground state and use this to provide us with
information about the system as we can indeed do for any other
gas. The general Hamiltonian that describes the system is given by
\begin{equation}
  \hat H=\sum_{i=1}^N\left(\frac{{\bf p}_i^2}{2m}+V_{ext}({\bf r}_i)\right) + \frac{1}{2}\sum_{i=1}^{N}\sum_{j\neq i}^N V\left(|{\bf r}_i-{\bf r}_j|\right),
  \label{eq:hamiltonian}
\end{equation}
where the first term on the right-hand-side is the kinetic energy
energy of the $i$-th particle, the following term represents the
external effects introduced by the trapping potential $V_{ext}$, and
the final term represents the interactions between the $N$
particles. The ground state corresponds to the minimum energy and thus
we can find it by minimising it. In that respect,
it is therefore convenient to use the concept of the thermodynamic
potentials which are useful in determining the equilibrium state of a
system not in isolation and are usually introduced early on in
undergraduate courses on thermodynamics and thermal physics
\cite{Gasser_Book1995, Huang_Book87}. Using the free energy, we have
that we need to minimise $F=E-\mu N$, where $E$ is the energy and
$\mu$ is the chemical potential.

Given a Hamiltonian $\hat H$ and a wavefunction $\psi$, we can obtain
the energy as follows:
\begin{equation}
  E(\psi)=\frac{\bra{\psi}\hat H\ket\psi}{\braket{\psi}{\psi}},
  \label{eq:energy}
\end{equation}
and as such we can use this expression to minimise the free energy
$F$. In the condensate we have $N$ particles and we can thus associate a
wavefunction $\psi_i$ to every one of them. However, in order to
capture the essential aspects of the problem we 
resort to making a mean-field approximation. This means that for one
particle, all the rest have the same status as they all are in the
same independent state $\ket{\psi}$ and thus we can drop the labelling
of the wavefunctions. In that way we need to
minimise the free energy over a space of functions of the
type $\ket{\Psi}=\ket{\psi} \otimes \ket{\psi} \otimes \cdots \otimes \ket{\psi}$,
where $\otimes$ represents the tensor product and thus $\ket{\Psi}$ is
the $N$-particle tensor product wavefunction; we are considering the
following normalisation $\braket{\Psi}{\Psi}=1$. This approximation is
valid if the condensate is 
not very dense; otherwise, the interactions with the closer neighbours
would be much stronger than with the particles that are farther apart.

Our problem is thus reduced to
minimising $F(\Psi)=\bra{\Psi} \hat H \ket{\Psi} - \mu
\braket{\Psi}{\Psi}$. Let us now compute each of the terms involved in
this calculation.  For the kinetic energy term, we have that
\begin{eqnarray}
  \bra{\Psi} \sum_{i=1}^N \frac{{\bf p}^2}{2m} \ket{\Psi} &=& \sum_{i=1}^N \frac{\hbar^2}{2m} \int \nabla \psi^*({\bf r}_i) \nabla \psi({\bf r}_i) d{\bf r}_i,  \nonumber\\
  &=& N \frac{\hbar^2}{2m} \int |\nabla \psi({\bf r})|^2 d{\bf r}, \nonumber\\
  &=& -N \frac{\hbar^2}{2m} \int \psi^*({\bf r}) \nabla^2 \psi({\bf r}) d{\bf r},
  \label{eq:kinetic_energy}
\end{eqnarray}
where as stated above $\ket{\Psi}$ is the $N$-particle tensor product
wavefunction and $\psi({\bf r})$ is the single particle wavefunction;
we have used Green's identity for obtaining the last line in equation
(\ref{eq:kinetic_energy}) \cite{Kaplan_Book1991}. The potential term
can readily be written as 
\begin{equation}
  \bra{\Psi} \sum_{i=1}^N V_{ext}({\bf r}_i) \ket{\Psi} = N \int \psi^*({\bf r}) V_{ext} \psi({\bf r}) d{\bf r}.
  \label{eq:potential}
\end{equation}
For the interaction term we have that
\begin{eqnarray}
  \bra{\Psi} \frac{1}{2} \sum_{i=1}^{N}\sum_{j\neq i}^N V\left(|{\bf r}_i-{\bf r}_j|\right) \ket{\Psi} \nonumber \\
  = \frac{1}{2}\sum_{i=1}^N \sum_{j \neq i}^N \int d{\bf r}_i \int \psi^*({\bf r}_i) \psi^*({\bf r}_j) V\left(|{\bf r}_i-{\bf r}_j|\right) \psi({\bf r}_i) \psi({\bf r}_j) d{\bf r}_j, \nonumber\\
  = \frac{N(N-1)}{2}\int d{\bf r} \int \psi^*({\bf r})\psi({\bf r}') V\left(|{\bf r}-{\bf r}'|\right) \psi({\bf r}) \psi({\bf r}') d{\bf r}'. 
  \label{eq:interaction}
\end{eqnarray}
Finally, for the last term in the free energy we have that
\begin{equation}
  \mu \braket{\Psi}{\Psi} = \mu \left( \int \psi^*({\bf r}) \psi({\bf r}) d{\bf r} \right)^N,
  \label{eq:chemical_potential}
\end{equation}
where we have written the expression in order to facilitate the
rest of our calculations.

Given the expressions above, we now need to minimise them. This can be
done following a procedure based on calculus of variations
\cite{ewing85,fox10}. In other words, we will consider a small
variation in the wavefunction $\psi({\bf r})$, but instead of varying
its real and imaginary parts, we take $\psi$ and $\psi^*$ as
independent variables. In that way, we can readily obtain the
functional derivatives $\frac{\delta\ldots}{\delta\psi^*}$ for
expressions (\ref{eq:kinetic_energy}) and (\ref{eq:potential}). In the
case of the interaction term, expression (\ref{eq:interaction}), we
have contributions from each of the two $\psi^*$, but the $\bf r$ can
be permuted, which results in the following expression
\begin{eqnarray}
   \frac{\delta}{\delta \psi^*} \bra{\Psi} \frac{1}{2}
   \sum_{i=1}^{N}\sum_{j\neq i}^N V\left(|{\bf r}_i-{\bf r}_j|\right)
   \ket{\Psi} \nonumber \\
 = N(N-1)\int \delta\psi^* ({\bf r})\left(\int |\psi({\bf r})|^2
   V(|{\bf r}-{\bf r}'|) d {\bf r}' \right)\psi ({\bf r}) d {\bf r} .
\label{eq:interaction_variation}
\end{eqnarray}
Similarly, for the chemical potential we have that
\begin{eqnarray}
  \frac{\delta \braket{\Psi}{\Psi}}{\delta \psi^*} &=& N\left(\int
    \psi^*({\bf r}) \psi({\bf r}) d {\bf r}\right)^{N-1} \int
  \delta\psi^*({\bf r}) \psi({\bf r}) d {\bf r}\nonumber\\
  &=& N\int \delta
  \psi^*({\bf r}) \psi({\bf r}) d{\bf r}.
\label{eq:mu_variation}  
\end{eqnarray}
Putting together all the different terms for the free energy $F$, we
have that the variation is given by:
\begin{eqnarray}
  \frac{\delta F}{\delta\psi^*} &=& 0 = N\int
  \left[-\frac{\hbar^2}{2m}\nabla^2 \psi({\bf r}) + V_{ext}(\bf
    r)\psi({\bf r}) \right.\nonumber \\
   &&\left.+ (N-1) \left(\int |\psi({\bf r})|^2
   V(|{\bf r}-{\bf r}'|) d {\bf r}' \right) \psi ({\bf r}) - \mu
 \psi(\bf r) \right] \delta 
\psi^*({\bf r}) d{\bf   r},
\label{eq:functional_F}
\end{eqnarray}
and therefore the quantity inside the square brackets in the above
expression must vanish.  It is quite common to choose an interaction
potential such that $V(|{\bf r}-{\bf r}'|)=\frac{4\pi\hbar^2}{m} a
\delta({\bf r}-{\bf r}')$ where $a$ is the $s$-wave scattering length
and using the  approximation that $N-1\simeq N$ we end up with 
\begin{equation}
  \label{eq:GPE_variation}
  -\frac{\hbar^2}{2m}\nabla^2 \psi({\bf r}) + V_{ext}({\bf
    r})\psi({\bf r}) +N \frac{4\pi\hbar^2}{m} a |\psi({\bf r})|^2\psi({\bf r})=\mu \psi({\bf
    r}),
\end{equation}
which is the time-independent Gross-Pitaevskii equation. The
scattering length $a$ measures the intensity of 
the interactions between the bosons. Its sign indicates whether the interactions are
attractive ($a<0$) or repulsive ($a>0$); we will later discuss the
importance of this distinction. In this way, the minimisation of
equation (\ref{eq:energy}) corresponds to the minimisation of the free
energy $F=E-\mu N$, which is a well-known result of statistical physics.

\subsection{Some comments about the GPE}
\label{subsec:commentsGPE}
As we mentioned above, the function $\psi$ is a complex valued
function. In that respect, following the standard interpretation in
quantum mechanics, its square modulus corresponds to a probability
density; given the
normalisation condition $\braket{\Psi}{\Psi}=1$, this quantity is the
probability density of finding a boson. We can equally choose the
normalisation such that
$\braket{\Psi}{\Psi}=N$, in this case the square modulus represents the
number density of the fluid.

The phase does not have an intrinsic meaning but its fluctuations
do. Let us write the wavefunction as $\psi=\sqrt{\rho}\exp({i\theta})$
and as such the probability current density can be
expressed as:
\begin{equation}
  {\bf j}=\rho{\bf v}=\frac{i\hbar}{2m}\left(\psi \nabla\psi^*-\psi^*\nabla\psi\right),
  \label{eq:probability_current}  
\end{equation}
where $\rho$ is the number density and 
\begin{equation}
  {\bf v}=\frac{\hbar}{m}\nabla \theta,
  \label{eq:velocity}  
\end{equation}
is the velocity of the bosons. We will come back to this point in
Section \ref{sec:hydrodynamic} where we present the hydrodynamic form
of the GPE. 

The external potential $V_{ext}$ allows us to model the action of the
external world on the condensate, in particular of a trap. Most of the
time, the bosons are confined in a small region of space (by optical
and/or magnetic means). This situation is simulated with the aid of a
trapping potential with a minimum in the central part. In general any
(regular) potential with a minimum can be locally approximated by a harmonic
potential, and such that the condensate remains in the part where the
quadratic term is dominant. Therefore,
\begin{equation}
  V_{ext}({\bf r})=\frac{1}{2}m\omega^2 r^2,
\end{equation}
where $m$ is the mass of the bosons, and $\omega$ corresponds to the
frequency of the trap.

In order to provide an idea of the orders of magnitude involved, we
summarise in Table \ref{tab:experimental} some data taken from real experiments
\cite{huepe,guery}.
\begin{table}[htbp]
\begin{center}
\begin{tabular}{|c||c|c|}\hline
  Quantity & Attractive condensate: $^7$Li & Repulsive condensate
  $^{87}$Rb\\ \hline\hline 
  $m$ & 1.16$\times 10^{-26}$ kg & 1.45 $\times 10^{-25}$ kg \\ \hline 
  $a$ & -23.3 $a_0$ & 109 $a_0$ \\ \hline
  $\omega$ & 908.41 rad$\cdot$s$^{-1}$ & 674.20 rad$\cdot$s$^{-1}$\\
  \hline 
  Typical size $r_0$ & 3.16 $\mu$m & 1.04 $\mu$m \\ \hline 
  $N$ & $\leq 1400$ & 100 000 \\ \hline
\end{tabular}
\caption{Data related to experimental realisations of Bose-Einstein
  condensation.\\ In the table, $a_0=0.529\times 10^{-10}$m is the Bohr
  radius.}
\label{tab:experimental}
\end{center}
\end{table}

In the case of rubidium atoms, if the bosons did not interact, the
ground state of the trapping potential on its own would be given by
$\psi({\bf r})\propto {\rm e}^{-(r/r_0)^2}$, where
$r_0\simeq\sqrt{\frac{\hbar}{m\omega}}\simeq 1.04\mu$m. However, in
the condensates obtained experimentally, the form of the ground state
is parabolic. If the condensates are relatively dense, the kinetic
energy would be negligible compared to the rest of the terms, and
therefore we have that
\begin{equation}
  \psi({\bf r})=\sqrt{\frac{\mu-m\omega^2r^2/2}{gN}},
\end{equation}
where $g=\frac{4\pi\hbar^2}{m}a$. From the above expression we can
see that the condensate has a parabolic form. The physics behind the
idea of neglecting the kinetic energy term is that the energy to add a
particle at any point in the cloud is the same everywhere. This energy
is given by the sum of the external 
potential $V_{ext}({\bf r})$ and an interaction contribution,
i.e. $gN|\psi({\bf r})|^2$ . This should be equal to the chemical
potential of a uniform gas having a density equal to the local density
$\rho=|\psi({\bf r})|^2$ \cite{Pethick_Book02}. In this context, this approximation is
referred to as the Thomas-Fermi
approximation \cite{Dalfovo99, Baym1996} due to its similarity to
the Thomas-Fermi 
approximation used for fermions \cite{greiner1998quantum}. 

\section{The time-dependent GPE}
\label{sec:timedepGPE}
The GPE that we have already established is related to the stationary
state of the condensate. Experimentally, one puts the atoms in
favourable conditions for the formation of a condensate, and waits for
the bosons to rearrange in a form such that the GPE is valid. However,
one might be interested to see what happens when the environment of the
condensate evolves or when the bosonic cloud is unstable. In order to
do that we need to 
establish a mean-field equation that will govern the evolution of
bosons in the minimum energy level.

According to quantum mechanics, bosons should obey the Schr\"odinger
equation $i\hbar\partial_t\Psi=\hat H\Psi$, where $\hat H$ is the
Hamiltonian, analogous to the one given by equation
(\ref{eq:hamiltonian}). In a similar way, we can consider that $\psi$
(describing the state of the bosons) satisfies a principle of minimum
action: the action $\int_{t_1}^{t_2}{\rm d}t\int{\rm d}^3{\bf r}L$,
where $L$ is the Lagrangian density, must be stationary. The Lagrangian density is
defined as
\begin{equation}
  L=i\frac{\hbar}{2}\left(\Psi^*\partial_t\Psi-\Psi\partial_t\Psi^*\right)-
  \frac{\hbar^2}{2m}(\nabla\Psi^*)\cdot(\nabla\Psi)-V_{ext}\Psi^*\Psi,
\end{equation}
for a single particle. Once we have written the Lagrangian density for our
problem, we have to simplify it using functions of the form
$\ket{\Psi}=\ket{\psi}\otimes\ket{\psi}\otimes\cdots\otimes\ket{\psi}$ that give
a stationary action, and thus will finally lead to equation (\ref{eq:GPE}).

It can be verified that equation (\ref{eq:GPE_variation}), which does
not include the temporal derivative,
is a particular case of the time-dependent GPE. We can therefore say
that the solutions of equation
(\ref{eq:GPE_variation}) correspond to stationary states of the gas. 

\subsection{Further remarks}
\label{subsec:tools}
Let $\tilde \psi$ be a solution of the GPE obtained using the
normalisation condition that $\braket{\tilde \psi}{\tilde
  \psi}=N(\tilde \psi)$ (refer to Section
\ref{subsec:commentsGPE}). If we multiply equation 
(\ref{eq:GPE}) by $\tilde\psi^*$, we get
\begin{equation}
  i\hbar\frac{\partial \tilde\psi}{\partial
    t}\tilde\psi^*=-\frac{\hbar^2}{2m}\nabla^2 \tilde\psi \tilde\psi^* +
  V_{ext}|\tilde\psi|^2+g|\tilde\psi|^4-\mu|\tilde\psi|^2.    
\end{equation}
Writing a similar equation for $\tilde\psi^*$ and adding them together we
obtain
\begin{equation}
  2i\frac{\partial}{\partial t}N(\tilde\psi) = \int{\rm d}^3{\bf r}
  \left[\frac{\hbar^2}{2m}|\nabla \tilde\psi|^2 +
    \frac{m\omega^2}{2}|{\bf r}|^2|\tilde\psi|^2+\frac{g}{2}|\tilde\psi|^4\right]-\mu N(\tilde\psi).
  \label{eq:multenergy}
\end{equation}
 The left-hand side of
(\ref{eq:multenergy}) is an imaginary quantity ($N(\tilde\psi) \in {\bf
  \mathbb{R}}$), whereas the right-hand side is real, with $N(\tilde\psi)$ being
constant and $\mu$ is fixed from the beginning, therefore the quantity
\begin{equation}
  E=\int{\rm d}^3{\bf
    r}\left[\frac{\hbar^2}{2m}|\nabla\tilde\psi|^2+\frac{m\omega^2}{2}|{\bf
      r}|^2|\tilde\psi|^2+\frac{g}{2}|\tilde\psi|^4\right] 
  \label{eq:Efunctional}
\end{equation}
must be conserved; it corresponds then to the total energy of the
bosonic gas. In this way, a solution of the GPE has two conserved
quantities, the number of particles $N$ and the energy $E$ of the
system. The existence of
these two quantities tells us some things about the 
eventual divergence of the time-dependent solutions: the norm of
$\tilde\psi$ is bounded, but the norm of $\nabla^2 \tilde\psi$ will diverge. The
eventual explosion of the solutions will be translated by the
appearance of peaks with infinite slope.

\subsection{Hydrodynamic form of the GPE}
\label{sec:hydrodynamic}
In general, we can use the time-dependent GPE as shown in equation
(\ref{eq:GPE}), nonetheless an equivalent set of equations can help us
shed light on other properties of the condensate. As pointed out in
Section \ref{subsec:commentsGPE}, for a solution of the form
$\psi=\sqrt{\rho}\exp(i\theta)$, one can use the density
$\rho$ and the gradient of its phase to describe the system.

With that in mind, let us derive a continuity equation by multiplying
the time-dependent GPE shown in equation (\ref{eq:GPE}) by $\psi^*$
and subtracting the
complex conjugate of the resulting equation. We end up with the
following expression:
\begin{equation}
  \label{eq:continuity}
  \frac{\partial |\psi|^2}{\partial
    t}+\nabla\cdot\left[\frac{i\hbar}{2m}\left(\psi\nabla\psi^*-\psi^*\nabla\psi \right)\right]=0.
\end{equation}
The equation above has the form of a continuity equation for the
particle density $\rho$ and can be expressed as follows:
\begin{equation}
  \label{eq:continuity_rho1}
  \frac{\partial \rho}{\partial t}+\nabla\cdot(\rho{\bf v})=0,
\end{equation}
where $\bf v$ is the velocity of the condensate and it is defined by:
\begin{equation}
  \label{eq:velocity1}
  {\bf v}=\frac{i\hbar}{2m}\frac{\left(\psi\nabla\psi^*-\psi^*\nabla\psi \right)}{|\psi|^2}=\frac{\hbar}{m}\nabla\theta.
\end{equation}
Equation (\ref{eq:velocity1}) tells is that the motion of the
condensate corresponds to that of a flow, since the velocity
corresponds to the gradient of a scalar quantity. If the phase
$\theta$ is not singular we can immediately see that the motion of the
condensate is irrotational:
\begin{equation}
  \label{eq:irrotational}
  \nabla\times{\bf v}=\frac{\hbar}{m}\nabla\times\nabla\theta=0.
\end{equation}
Please note that the result above is only valid when $\theta$ is not
singular; this would not be the case at the core of a vortex line. In
that case, the single-valuedness of the condensate wavefunction
implies that around a closed contour its change in phase,
$\Delta\theta$, must be a
multiple of $2\pi$:
\begin{equation}
  \label{eq:closed_contour}
  \Delta\theta=\oint\nabla\theta\cdot d{\bf l}=2\pi l,
\end{equation}
where $l$ is an integer. In that way, the circulation $\Gamma$ around
a closed contour is given by:
\begin{equation}
  \label{eq:circulation}
  \Gamma=\oint {\bf v}\cdot d{\bf l}=\frac{\hbar}{m}2\pi l.
\end{equation}
In the case of having a singularity in the phase, equation
(\ref{eq:circulation}) shows that the circulation is 
quantised. 

So far, equation (\ref{eq:continuity_rho1}) provides us with the
equation of motion for $\rho$. Let us obtain an equation of motion for
the phase $\theta$. Substituting a wavefunction of the form 
$\psi=\sqrt{\rho}\exp{(i\theta)}$ into equation
(\ref{eq:GPE}), we can then identify the real and imaginary parts of the
equation. From this procedure it can be shown that the imaginary part
of the equation results in equation (\ref{eq:continuity_rho1}),
whereas the real part is given by:
\begin{equation}
  \label{eq:motion_theta}
  -\hbar \frac{\partial \theta}{\partial
    t}=-\frac{\hbar}{2m\sqrt{\rho}} \nabla
  \sqrt{\rho}+\frac{1}{2}m{\bf v}^2+V_{ext}+g\rho. 
\end{equation}
Taking the gradient of equation (\ref{eq:motion_theta}) we have that
\begin{equation}
  m\frac{\partial {\bf v}}{\partial t}+\frac{1}{2}m\nabla{\bf v}^2=
  \frac{\hbar^2}{2m}\nabla
  \left(\frac{1}{\sqrt{\rho}}\nabla^2\sqrt{\rho}\right)-\nabla
  V_{ext}-g\nabla\rho,
  \label{eq:hydro2}
\end{equation}
where we have used equation (\ref{eq:velocity1}) to write the expression in terms
of the velocity $\bf v$. For an irrotational flow, given that
$\frac{1}{2}\nabla{\bf v}^2=({\bf v}\cdot\nabla){\bf v}+{\bf v}\times
(\nabla\times {\bf v})$, we obtain an equation analogous to the 
Navier-Stokes equation without viscosity or Euler equation:
\begin{equation}
\label{eq:Euler}
  m\frac{\partial {\bf v}}{\partial t}+m({\bf v}\cdot\nabla){\bf v}=
  \frac{\hbar^2}{2m}\nabla
  \left(\frac{1}{\sqrt{\rho}}\nabla^2\sqrt{\rho}\right)-\nabla
  V_{ext}-g\nabla\rho.  
\end{equation}

Equations (\ref{eq:continuity_rho1}) and (\ref{eq:hydro2}) are exactly
equivalent to the GPE, and thus it is possible to use the latter to
investigate the behaviour of the fluid. The GPE
is, from a mathematical point of view, equivalent to the equations
obtained from fluid mechanics. In the case of a condensate with
non-singular phase, the BEC behaves like a
fluid without viscosity and with irrotational flux.

\section{Application of the GPE to a cloud of bosons in one dimension}
In this section, we demonstrate the application of the GPE to a relatively simple case: a
bosonic gas in 1 dimension in a uniform potential. Such a gas can be
obtained by getting a condensate in an elongated trap. If the
temperature is low enough, the thermal agitation energy $k_BT$ will
not exceed the energy threshold of the trap ($\hbar\omega/2$) in the
directions perpendicular to the axis of the trap and they could be
neglected. The case of the elongated trap corresponds to looking for a
solution over all space, and we are particularly interested in the
phenomena happening in the centre of the trap, where the density is
relatively uniform or, in other words, where the condensate behaves as
a fluid. 

In order to tackle this problem, let us consider a condensate confined
by a box with infinitely hard walls. At the wall the wavefunction must
vanish; and inside it, the condensate density approaches its bulk. Let
us assume that the potential vanishes for $x\le 0$ and infinite for
$x<0$. In the $y$ and $z$ directions the wavefunction is uniform, and
thus equation (\ref{eq:GPE_variation}) takes the following form:
\begin{equation}
  \frac{-\hbar^2}{2m}\frac{{\rm d}^2\psi}{{\rm d}x^2}+U_0|\psi|^2\psi=\mu\psi.
  \label{eq:1D}
\end{equation}
where we have written $U_0=N\frac{4\pi\hbar^2}{m}a$. In the case of a uniform gas, it can be seen that the chemical potential is given by
$\mu=U_0|\psi_0|^2$, where $\psi_0$ is the wavefunction far from the
wall. The equation then becomes
\begin{equation}
  \label{eq:1D_Approx}
  \frac{-\hbar^2}{2m}\frac{{\rm d}^2\psi}{{\rm
      d}x^2}=U_0\left(|\psi|^2-|\psi_0|^2 \right)\psi. 
\end{equation}
When we consider repulsive interactions, i.e. $U_0>0$, this equation,
subject to the boundary conditions $\psi(0)=0$ and 
$\psi(\infty)=\psi_0$, has the solution
\begin{equation}
  \label{eq:darksoliton}
  \psi(x)=\psi_0\tanh\left(\frac{x}{\xi\sqrt{2}} \right),
\end{equation}
where $\xi^2=\frac{\hbar^2}{2m|\psi_0|^2U_0}$; this quantity describes the
distance over which the wavefunction tends to its bulk value when
subjected to local perturbations, and it is therefore referred to as
the healing length \cite{Leggett01}.

In the case of attractive interactions ($U_0=-|U_0|$) the solution is
given by 
\begin{equation}
  \label{eq:brightsoliton}
  \psi(x)=\sqrt{2}\psi_0 {\rm sech}\left(\frac{x}{\xi} \right).
\end{equation}
In contrast to the case with repulsive interactions, the density
vanishes at large distances from the centre. Since $U_0<0$, the
interaction energy decreases with increasing density, and thus the
condensate collapses \cite{Ruprecht}. Taking into account the trapping potential, it is
possible to obtain metastable condensates as has been shown by Bradley
et al. \cite{Bradley}.

\subsection{Excitations}
\label{subsec:sinexcit}
 It is possible to linearise the GPE \cite{cohen} or its hydrodynamic form
\cite{nore}. In the latter case, we set $\rho=\rho_0+\delta\rho$,
where $\rho_o=\frac{\mu-V}{a}$, and ${\bf v}=\delta{\bf v}$ and
substitute into equations (\ref{eq:continuity_rho1}) and (\ref{eq:hydro2}). In
the case of equation (\ref{eq:hydro2}), written in a normalised
formalism and keeping first order terms only, we have that
\begin{equation}
  \frac{\partial\delta{\bf v}}{\partial t}+a\nabla\delta\rho-
  \frac{1}{4\rho_0}\nabla\nabla^2\delta\rho=0. 
\end{equation}
Taking the divergence of this equation:
\begin{equation}
  \frac{\partial\nabla\cdot\delta{\bf v}}{\partial t}+ 
  a\nabla^2\delta\rho-\frac{1}{4\rho_0}\nabla^2\nabla^2\delta\rho=0.
\end{equation}
From equation (\ref{eq:continuity_rho1}), keeping only first order terms, we
have
\begin{equation}
  \frac{\partial\delta\rho}{\partial t}+\rho_0\nabla\cdot\delta{\bf v}=0.
\end{equation}
Finally, combining these two expressions, we get the following
equation
\begin{equation}
  -\frac{\partial^2 \delta\rho}{\partial
    t^2} + a\rho_0\nabla^2\delta\rho -
  \frac{1}{4}\nabla^2\nabla^2\delta\rho=0, 
\end{equation}
which can be seen as the equation of propagation of sound waves in the
fluid. It corresponds to the dispersion relation, obtained by taking
the Fourier transform of $\rho$, namely
\begin{equation}
  \omega^2=\mu k^2+\frac{k^4}{4},
\end{equation}
with $\mu<0$ and $a<0$. We can therefore verify the existence of a
critical value for the wavenumber $k$: if $k<\sqrt{-2\mu}$, then
$\omega^2<0$ and the periodic perturbations of the wavenumber tend to
diverge from the stable solution (there is an explosion). If on the
contrary, $k\geq\sqrt{-2\mu}$, the perturbed solutions oscillate
around the constant solution.

\section{Conclusion}
In this paper we have derived the Gross-Pitaevskii equation using
concepts covered in many advanced undergraduate courses. The
discussion was motivated from the point of view of the behaviour of a
gas as explained from results of statistical mechanics. We have seen
how a many-body problem in quantum mechanics can be 
solved using suitable approximations and using a nonlinear partial
differential equation which gives relatively good results. Even if
the Gross-Pitaevskii equation seems somewhat simple, it can be
applied to obtain information about a great number of physical systems
such as Bose-Einstein condensates, superfluids or light phenomena. In
particular, in this paper we have made use of the GPE to model the
ground state of a Bose-Einstein condensate as well as starting to address its
excitations in a low-temperature limit. We expect that this paper
encourages discussions of modern developments in physics within the
context of advanced coursed in statistical mechanics, mathematical
physics or physical modelling.

\section*{Acknowledgements}
The author would like to thank Dr James Collett for reading and
commenting on the manuscript.

\section*{References}
\bibliographystyle{unsrt} \bibliography{GPE}

\end{document}